\documentclass[aps,prl,twocolumn,showpacs]{revtex4-1}

\usepackage{graphicx,amsmath, hyperref,booktabs,multirow}

\def\toprule{\hline\hline}%
\def\botrule{\hline\hline}%

\begin{document}


\title{Investigating the Effectiveness of the Tutorials in Introductory Physics in Multiple Instructional Settings}
\author{
C.\ Slezak$^{1}$,
K.\ M.\ Koenig,
$^{2}$, 
R.\ J.\ Endorf
$^{2}$ and
G.\ A.\ Braun$^{3}$}

\affiliation{
$^{1}$Hillsdale College, Hillsdale, Michigan, 49242, USA \\
$^{2}$University of Cincinnati, Cincinnati, Ohio, 45221, USA \\
$^{3}$Xavier University, Cincinnati, Ohio 45207}

\date{\today}

\begin{abstract}
This paper examines the educational impact of the implementation of ''Changes in Energy and Momentum'' from the Tutorials in Introductory Physics in five different instructional settings.  
These settings include (1) a completely computer-based learning environment and (2) use of cooperative learning groups with varying levels of instructor support.  Pre- and post-tests provide 
evidence that a computer-based implementation falls significantly short of classroom implementations which involve both collaborative learning groups and interactions with a teaching assistance. 
Other findings provide insight into the importance of certain elements of instructor training and the appropriate use of the tutorial as an initial introduction to a new concept.
\end{abstract}
\maketitle

\section{Introduction}

Newly developed materials from the physics education research (PER) community are beginning to reform the way physics is being taught across the country with mixed success. The secondary 
implementation of some of these materials has been shown to be vastly successful in some studies \cite{tutorialsCU,mazur,redish,scaleup} while suffering from startup-complications and other difficulties in other 
studies \cite{studio,sabella,gender}. Faculty training workshops, instructor manuals, and literature resources are widely available, but are not always consulted or able to be implemented. The PER 
community continually facilitates the adoption of new materials and pedagogical approaches at secondary locations, but institutional challenges combined with institutional inertia may result in 
implementation contrary to those which were initially conceived. This necessitates further exploration as to the effectiveness of the utilized materials in a multitude of settings along with
improved assistance in implementation.

{\em The Tutorials in Introductory Physics}\cite{tutorials} is one such research-validated curriculum which has been adopted by many institutions in an attempt to improve student learning in introductory calculus-based 
physics courses. These inquiry-based materials are intended to aid in the transition towards reformed instruction by replacing traditional recitations, while not altering the common course 
structure of separate lecture, recitation, and lab. The tutorials contain a collection of worksheets that are designed such that students work in small collaborative learning groups and engage in 
conceptual activities that address common difficulties encountered in studying physics.  At various points during the activities student reasoning is checked by an instructor or teaching assistant.
The use of Socratic Dialogue, or directed questioning, during these checkpoints helps bring students to correct understanding.  The tutorials have been shown to significantly increase student 
conceptual understanding and scientific reasoning skills when compared to more traditional recitation settings\cite{tutorialsUC,emtutorials,mazur2}.  

Although improved student learning is a primary reason for the adoption of new instructional techniques, often these implementations are constrained by time and available resources. For example, 
a successful implementation of the {\em Tutorials in Introductory Physics} curriculum requires a small student-to-teacher ratio and has been shown to depend on highly trained instructional 
staff\cite{tutorialsUC}. These requirements may put a financial burden or pose staffing difficulties on already stretched departments. As a result, educational reforms might be implemented 
under less than ideal conditions, as when using inexperienced teaching assistants, or they might be implemented in ways not originally intended, such as through computer- or web-based instruction.

A move towards computer-based instruction has been the subject of multiple studies\cite{Podolefsky}. For example, conceptual gains in student understanding have been found in implementations that substitute 
computer simulations for actual laboratory equipment\cite{PHETlabs}, augmenting lecture components via web-based multimedia pre-lectures\cite{prelectures}, and Just-in-Time-Teaching strategies\cite{jitt}. However many other 
research-validated curricula have not been implemented in a computer-based format and this paper adds to the research literature regarding the use of such an environment with an activity from 
the {\em Tutorials in Introductory Physics}.

\section{Objectives and Methods}

This study addresses two research questions.  First, although the effectiveness of utilizing computer simulations in conjunction with the {\em Tutorials in Introductory Physics} has been previously
demonstrated \cite{PHETtut}, this study investigates the impact of a computer-based environment on student learning when the computer is used rather than a teaching assistant to check student understanding 
at designated checkpoints within the tutorials.  Second, it attempts to build upon our previously reported study which gauged the effectiveness of various levels of instructional implementation of 
the tutorial materials\cite{koenig}.

In our prior study, students completed the tutorial activity ''Changes in Energy and Momentum'' by participating in one of four styles of recitation including (1) a traditional lecture on the topic 
with answers to the tutorial activity provided by the instructor, (2) students working independently through the tutorial worksheets with brief written answers provided at checkpoints, (3) 
students working in cooperative learning groups with brief written answers provided at checkpoints, and (4) students working in cooperative learning groups with a highly trained teaching assistant 
facilitating checkpoints using Socratic dialogue (see Table~\ref{stylecomp}).  We were able to demonstrate the importance of the instructor's role when implementing the tutorials as significantly higher gains in 
student learning were only observed when the instructional setting included the use of Socratic dialogue with a highly trained teaching assistant.  Surprisingly, the use of cooperative learning 
groups and brief written answers provided at checkpoints did not result in improved student learning when compared to recitations that used only lecture or individual work.  As part of our current 
investigation we seek to further explore the relationship between the role of the instructor and student learning.

\subsection{Design}

In order to maintain a high level of parity with our previous study, the experimental setup was closely replicated. This study was also performed at the University of Cincinnati with students who 
were primarily first year engineering majors.  The structure of the course had not changed in the year between the two studies 
although some of the instructors or teaching assistants were different.  Four sections of calculus-based physics were offered and taught by four different instructors.  Students met three times 
per week for 50-minute lectures of roughly 120 students with one weekly 50-minute recitation of 25 students.  The {\em Tutorials in Introductory Physics} continued to be used in recitation which was 
taught jointly by a graduate and undergraduate teaching assistant. 

As before, participants in the subsequent study were drawn from the first quarter of our introductory calculus-based physics course.  A total of 200 students of the 465 students enrolled in the 
course volunteered to participate in the study by agreeing to complete one additional recitation which was not part of the regular course. Students received extra credit on the basis of their 
participation alone and the same amount of extra credit was offered to students who alternatively completed an additional written assignment. In order to accommodate individual student schedules, 
multiple participation times were provided during 4 consecutive days and students were randomly assigned to one of the 5 different experimental groups.

The tutorial topic chosen for this study, ''Changes in Energy and Momentum'', was the same as used in the prior study.  This tutorial is not traditionally covered in the recitations at our university,
but it addresses concepts students traditionally struggle with when asked to apply them: the work-energy theorem and the impulse-momentum theorem. In addition, the choice of this topic allows us 
to compare our results to previously reported investigations\cite{koenig,koenig2, tutorialsUW}.

The choices of tutorial implementations that comprise this study were designed to augment those used in our prior study in addition to providing information on the use of a computer-based 
environment. Table~\ref{stylecomp} portrays the styles of implementation employed in the two studies.  The original study found that only the use of a highly trained instructor implementing Socratic dialogue 
at checkpoints yielded significantly higher gains in student learning.  This suggests that the written materials do not have as much of an impact on student learning as the performance of the 
instructor. Although all teaching assistants go through the same weekly tutorial training, which is limited to the verification of content understanding with no pedagogical issues or strategies 
discussed due to time constraints, like others, we find large variations in the how the tutorials are implemented in the recitation classrooms\cite{TAbelieves}.  At designated checkpoints, some teaching assistants visually 
check students’ written answers and then through oral or written means indicate what the answers should have been.  
Some go into detail about the actual reasoning behind the correct answers while others do not.  Other teaching assistants make attempts to engage students in Socratic dialogue but aren't skilled 
enough to correctly identify and address incorrect student thinking. Unfortunately, each term we typically identify only one of a dozen teaching assistants who are able to implement the tutorials 
in a manner consistent with what we deem the ideal level.  That is, they are able to engage students in Socratic dialogue to probe their reasoning, correctly identify when a misconception exists, 
and lead students to correct reasoning through appropriately directed questions.  

In light of this, we set up the conditions for four of the five experimental groups in the second study to model the more subtle differences we have observed amongst our teaching assistants when implementing the 
tutorials in recitation.  That is, although one experimental group was provided with a computer-based environment in which students worked through the tutorials individually (Style 1), 
the other four experimental groups involved the use of cooperative learning groups with slightly different levels of teaching assistant interaction during checkpoints (Styles 2-5).  
\begin{center}
\begin{table*}[ht]
{\small
\hfill{}
\begin{tabular}{lcl}
\toprule
Implementations of Prior Study & Forum & Implementations of Current Study\\
\hline
Lecture & \multirow{2}{*}{Individual} & Computer-based environment (Style 1)\\
Brief written answer keys provided\\
\hline
Brief written answer keys provided & \multirow{4}{3cm}{\centering Cooperative Learning Groups} &Detailed written answer keys (Style 2)\\
Ideal implementation* & & Detailed oral answers (Style 3)\\
  &  &Limited Socratic Dialogue (Style 4)\\
& & Ideal Implementation* (Style 5)\\
\botrule
\multicolumn{3}{p{13.5cm}}{*These were designed to be identical implementations between the two studies although both were conducted by a different graduate teaching assistant.} \\
\end{tabular}}
\hfill{}
\caption{Styles of implementation of the two studies}
\label{stylecomp}
\end{table*}
\end{center}
In the present study, Style 1 was the only implementation which did not involve cooperative learning groups. Students completed the tutorials individually on a workstation in a computer lab. 
One of the authors was present in the lab for technical support – albeit none arose.  This author also provided a brief introduction on how to use the computer-based system.  The students worked 
through the entire tutorial activity in this interactive computer-based environment with all “text” from the tutorials displayed on the computer screen.  Scrap paper was provided so students could 
work out the answers to the problems on paper before transferring them to the system.  Questions required students to type in their responses in the form of equations, derivations and/or 
explanations. Although the majority of the questions were identical to those in the printed version of the tutorial, the questions that asked for graphical representations of vectors were 
rephrased and students chose from a list of possible vector directions rather than actually draw the vector.  

In all cases the students worked through the material in the same order as in the printed version and feedback was provided at the same points where all other styles had instructor checkpoints. 
At these checkpoints, the program would provide the students with dynamic feedback on submitted answers. Correct answers were reaffirmed with a short summary of the main concept and restatement 
of the correct answer and associated reasoning.  Incorrect answers were identified as such in the feedback and more detailed explanations were provided addressing the mistake made. In the second 
half of the tutorial the students were to engage in an experimental piece.  Similar to our previous study, rather than have the students actually set up and observe the trajectory of a 
ball as it rolls down a ramp, this activity was replaced in all styles with a pre-drawn path which was provided to the students in print form. 

The other four recitation styles in the current study mimicked what is typical in our recitation classroom settings.  That is, the students worked in cooperative learning groups of 3 or 4 
students under the guidance of one graduate and one undergraduate teaching assistant.  The differences between these four implementations were in the types of allowable interactions between 
teaching assistants and students.  The same graduate and undergraduate teaching assistants conducted these implementations in the study with the exception of Style 5 which was conducted 
solely by the graduate teaching assistant.  Both teaching assistants were experienced and had previously taught using the tutorials in recitation for approximately 2 years each.

In Style 2, students worked through the material in cooperative learning groups.  At designated checkpoints students were provided with a written answer key that indicated both the correct answer 
along with detailed explanations of the reasoning behind the answer. This approach is in contrast to one of the styles used in our previous study in which cooperative learning groups were used but 
only minimalistic written answers, without any form of explanations, were provided at checkpoints. In both implementations verbal interactions between the teaching assistants and students regarding 
tutorial content were not allowed.

In Style 3 the teaching assistants checked student answers at each checkpoint but in this case correct answers and the supporting reasoning were provided verbally.  This information was carefully 
scripted so it matched what was written on the student answer keys disseminated in Style 2. The teaching assistants were encouraged to review student answers for correctness and comment on them 
but were to not engage the students in directed questioning to lead students to correct understanding.  This style was set up to model what we often observe in our regular recitation classes 
facilitated by our graduate students.  That is, our regular teaching assistants typically check student answers and provide correct answers, when necessary, without attempting to find out where 
students are struggling nor do they attempt to help students work through incorrect ideas.  

Style 4 took this one step further by allowing the teaching assistants to implement limited Socratic dialogue with the focus of this conversation restricted to only those misconceptions and 
talking points identified in the tutorial instructor guide\cite{instguide} as the most common for this question.  This style was designed to model the classroom situation where we have witnessed teaching 
assistants not well versed in handling a variety of student misconceptions.  In these cases the teaching assistants either ignored or misinterpreted student misconceptions and the dialogue tended 
to center on only those misconceptions that were addressed in our weekly tutorial training meetings.   

And last, Style 5 allowed for full Socratic dialogue upon the discretion of the teaching assistant.  This style was designed to model an ideal implementation of the tutorials with the teaching 
assistant probing student reasoning, correctly identifying student difficulties, and subsequently leading students to correct reasoning through appropriately directed questions.  

Prior to conducting each implementation style, the graduate and undergraduate teaching assistants were carefully trained by one of the authors. In addition, at least one of the authors was present 
during each implementation with students to monitor that the session was conducted as designed.  No discrepancies were noted. For Styles 3-5, all verbal interactions between students and teaching 
assistants were recorded for later review if necessary.

\subsection{Evaluation}

\begin{figure}[htb]
\begin{center}
\leavevmode
\includegraphics{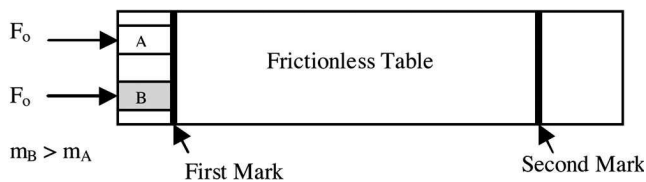}
\end{center}
\caption{Experimental setup used for all pre-test questions.}
\label{fig:q1}
\end{figure}

\begin{figure}[htb]
\begin{center}
\leavevmode
\includegraphics{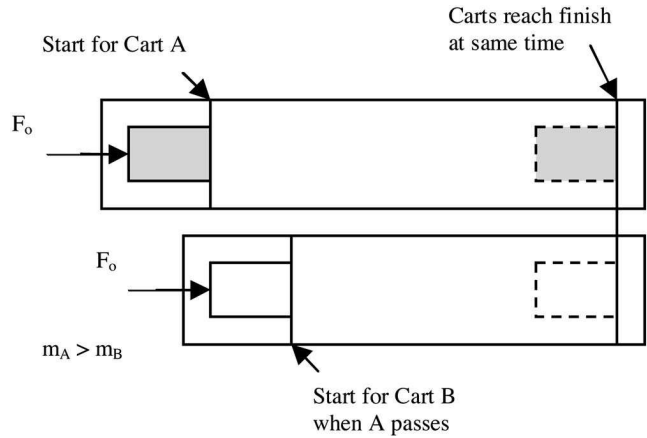}
\end{center}
\caption{Second experimental setup used for post-test questions.}
\label{fig:q2}
\end{figure}

The instructional impact of the implementation styles on students’ conceptual understanding was measured by administering the established pre- and post-test for this tutorial developed by members of the 
Physics Education Group at the University of Washington\cite{tutorialsUW,instguide}. The pretest considers the experimental setup shown in Figure~\ref{fig:q1}. Two carts of different masses
$m_B > m_A$ are pushed from rest by equal and constant forces $F_0$ between two depicted lines on a frictionless table. Students are asked to compare the kinetic energy and momentum of both masses 
as they cross the second mark. The same comparisons are asked for in the post-test which consists of two different scenarios as shown in Figure~\ref{fig:q2}. The first scenario is identical to 
that of the pre-test but instead of the carts being pushed equal distances, identical forces are applied over the same time interval $t_A = t_B$. The second scenario involves cart $A$ being 
accelerated with constant force $F_0$ from rest. At the instance it passes stationary cart $B$, an identical force is applied to cart $B$ and both carts reach the finish line at the same time. 

The students participating in each implementation style completed the pre- and post-tests in paper and pencil form.  All completed the pre-test immediately before the tutorial activity and the 
post-test immediately after.  Two of the authors graded both the pre- and post-tests and only student responses for each of the respective questions which had both the correct comparison and 
accompanied reasoning were considered correct.  This grading process was identical to our previous study. The inter-rater reliability in the evaluation of student answers was found to be greater 
than $99\%$.  Throughout this paper, unless otherwise noted, all statistical comparisons of mean scores between implementation styles employed a multiple/Post-Hoc ANOVA comparison utilizing a 
Least Significant Difference (LSD) test.

\section{Results and Discussion}

\subsection{Determining Homogeneity of Groups}

The pre-test results for each of the five tutorial implementations are show in Table~\ref{pretest}. A one-way ANOVA showed no significant difference between groups for the kinetic energy or momentum pretest 
question ($p=0.39$ and $p=0.60$ respectively) indicating students had similar prior understanding of the topics. The pre-test mean scores for the 4 experimental groups in our prior study ranged 
from $21\%$ to $29\%$ for kinetic energy and $0\%$ to $3\%$ for momentum. A one-way ANOVA comparing pre-test mean scores of all nine experimental groups across both studies showed no statistical differences between 
groups for both topics of kinetic energy ($p=0.15$) and momentum ($p=0.58$).  This was expected as the pre-test was administered at the same point in the calculus-based physics curriculum during 
both studies. Although the pre-test results for both the kinetic energy and momentum comparisons indicated low levels of student prior knowledge, the higher mean on the questions associated with 
kinetic energy may be attributed to the fact that at the time of this study, students had been exposed to the work-energy theorem but not the impulse-momentum theorem. 

\begin{table}
\caption{Student performance on two pretest questions.}
\label{pretest}
\begin{tabular}{lcc}
\toprule

  & Correct kinetic & Correct momentum \\
  Recitation style &  energy comparison & comparison \\
\hline
Style 1 (N=29) & 7\%     &  0\% \\
Style 2 (N=45) & 18\%     &  4\% \\
Style 3 (N=38) & 11\%     &  0\% \\
Style 4 (N=41) & 22\%     &  2\% \\
Style 5 (N=47) & 13\%     &  2\% \\
\botrule
\end{tabular}
\end{table}

As an additional indicator of homogeneity of our experimental groups in this study, we considered the students’ exam averages in their calculus-based physics course.  All students took identical 
block exams as part of the course and no significant differences were found among these averages ($p=0.14$). Although this comparison was also done within our prior study, we were not able to 
compare student exam performance across studies as the block exams were not identical. 

\subsection{Impact of Computer-based Instruction}

The results for student performance on the post-test questions associated with kinetic energy and momentum are shown in Table~\ref{posttestcomp}.  Data from both of our studies are included for comparison 
purposes. The order of the styles listed in the table was chosen to reflect our expectations in student performance for each implementation. That is, we expected implementations that involve 
students working individually through the tutorials to have lower performance than those working in cooperative learning groups.  For implementations involving group work, we listed them based on 
increases in the amount of instructional support provided by the teaching assistants. 

\begin{table}
\caption{Student performance on post-test for various implementation styles.}
\label{posttestcomp}
\begin{tabular}{lcc}
\toprule

  & Correct kinetic & Correct momentum \\
  Recitation style &  energy comparison & comparison \\
\hline
Style 1 (N=29) & 16\%     &  7\% \\
Individuals* (N=76) & 19\%     &  14\% \\
Lecture* (N=75) & 26\%     &  20\% \\
Groups* (N=58) & 12\%     &  17\% \\
Style 2 (N=45) & 22\%     &  32\% \\
Style 3 (N=38) & 32\%     &  36\% \\
Style 4 (N=41) & 43\%     &  30\% \\
Style 5 (N=47) & 31\%     &  31\% \\
Ideal* (N=63) & 50\%     &  41\% \\
\botrule
\multicolumn{3}{p{8cm}}{* indicates results from our previous study\cite{koenig}}
\end{tabular}
\end{table}

One of the motivations for this study was to determine the level of effectiveness of a computer-based implementation of the tutorials.  As more and more courses move to online or computer-based 
instruction in an effort to reduce costs or provide more consistent implementation, we determined this to be an important approach to investigate.  In our previous study significant gains were 
observed only in what we deemed the ideal implementation of the tutorials, i.e. students working in cooperative groups under the guidance of a highly trained teaching assistant who was able to 
effectively engage students in Socratic dialogue at checkpoints. This finding surprised us as we expected the other style of implementation that involved students working in groups to have higher 
gains in student learning than those in which they worked individually.  We now believe this unexpected finding may be due to the type of feedback we provided students at checkpoints. That is, 
students were only provided with brief answers as shown in Figure~\ref{answerkey} for one of the tutorial checkpoints. We had hypothesized that a computer-based implementation with more robust onscreen feedback 
during designated checkpoints may be more effective. However, what we found was that the post-test scores of these students were the lowest of all implementations in the two studies with only 
$16\%$ and $7\%$ of students having both correct answers and reasoning for kinetic energy and momentum, respectively.

\begin{table}
\caption{Representative sample of tutorial question and corresponding answers key from the two studies.}
\label{answerkey}
\begin{tabular}{p{8cm}}
\toprule
{\bf Question I.D-1:} How does the net work done on cart $A$ ($W_{net,A}$) compare to the net work done on cart $B$ ($W_{net,B}$)? Explain.\\
\hline
{\bf Answer provided in previous study:} the net works are equal\\
\hline
{\bf Answer provided in this study:} Recalling the definition of work $W_{net} = \vec{F}_{net} \cdot \Delta \vec{x}$ only depends on the length
of the path in direction of the net force. Since in this setup both carts experience the same
net force over the same distance traveled they also have the same work done on them:
$W_{net,A} = W_{net,B}$\\
\botrule
\end{tabular}
\end{table}

In comparing the computer-based implementation to the implementation in our previous study which involved students working independently through paper versions of the tutorials with only brief 
written answers provided at checkpoints, we found no significant differences in student understanding of kinetic energy or momentum ($p=0.64$ and $p=0.40$, respectively). Although we expected the 
lack of cooperative learning groups to impact student gains in understanding, we expected the computer-based implementation to yield higher post-test scores due to more complete answers being 
provided at checkpoints.  However, we noted during the actual implementation of both styles that students in the computer-based environment spent on average $34\pm 1$ (actual times range from 18 to 52 minutes) 
minutes completing the tutorial activity whereas students completing the paper version averaged approximately 50 minutes (our records were not complete to provide an exact average).   
This difference in time on task may explain why the computer-based implementation yielded such low post-test scores. 

In comparing the computer-based implementation to all other styles in the two studies, the analyses became more complex as the patterns in student understanding of momentum on the post-test did 
not match those for kinetic energy.  That is, students of the computer-based implementation had statistically lower ($p<0.05$)  post-test scores for the momentum questions when compared to any other 
style in the present study as well as the ideal implementation of the tutorials in the previous study. But while student post-test scores for computer-based instruction are the lowest of any style 
considered, for the topic of kinetic energy the post-test scores of the computer-based implementation were only statistically lower than Style 4 and the ideal implementation within the previous 
study.  This was somewhat surprising because the comparisons here involved students working individually (i.e.~computer-based implementation) and students working in cooperative learning groups.  
We expected statistically higher gains in both topics for the implementations involving cooperative learning.  These differences in patterns of student performance on the post-test for the kinetic 
energy and momentum questions may be due to student prior knowledge of the two topics.  From the pre-test scores it is evident that students had prior knowledge of the work-energy theorem, but few 
students demonstrated knowledge of the impulse-momentum theorem. This suggests that the level of instructional support is more important when students are more familiar with the material.  This 
will be further addressed in the next section.

\subsection{Impact of Instructor}

As a second motivation for this study, we wanted to identify what teaching practices support more effective learning in recitations that use the tutorials.  Because some of the styles in both 
studies are similar, we expected comparable post-test results. For example, in our prior study one implementation style involved the use of cooperative learning groups with brief written 
answers provided at checkpoints.  Style 2 in the present study was similar except the written answers included additional detail on the reasoning behind the correct answers. Figure~\ref{answerkey} 
shows the differences in provided student feedback between the two studies for one checkpoint question in the tutorial. Although Style 2 demonstrated increases in student post-test scores for both the kinetic energy 
and momentum questions, statistical significance was only found for the momentum questions ($p=0.42$ and $p<0.05$ respectively).  

\begin{figure}[htb]
\centering
\includegraphics[width=1.0\columnwidth]{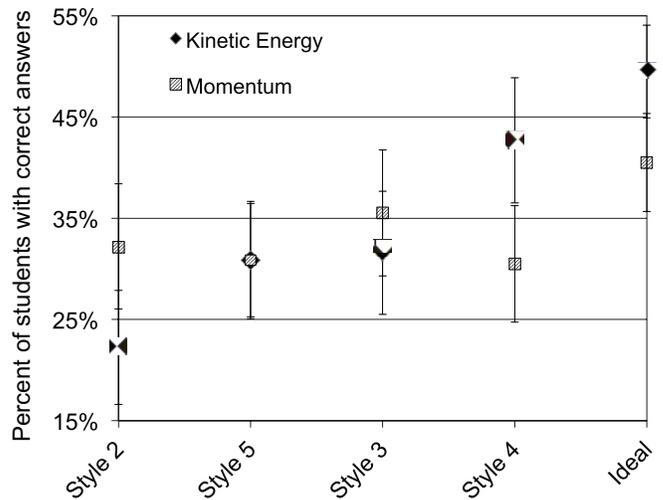}
\caption{Student post-test performance for recitation styles involving cooperative learning groups and verbal interactions with teaching assistants. }
\label{fig:chart}
\end{figure}

In order to more closely examine the impact of the instructor, we compared all styles in both studies that included both the use of cooperative learning groups as well as some level of verbal 
interaction with the teaching assistant.  This consisted of Styles 2-5 in the current study along with the ideal implementation in the prior study.  Figure~\ref{fig:chart} displays the mean scores 
for each of these implementations relative to student performance on both the kinetic energy and momentum questions.

Although the ideal implementation of the prior study appears to stand out in Figure~\ref{fig:chart} in terms of student performance on the post-test for the momentum questions, no statistical difference in 
student performance is found ($p=0.65$) amongst all five displayed styles.  As previously indicated, the completion of this tutorial activity was the first time most students were exposed to the 
momentum-impulse theorem as is evidenced by the low pre-test scores.  Because knowledge of this theorem is necessary when answering the momentum post-test questions, it appears that the tutorial 
material itself, rather than the quality of instructor support, most impacts student understanding.  In fact, the mean score of all students in Figure~\ref{fig:chart} is $34\%$, indicating that 
even after completing this tutorial activity, only one third of the students were able to provide correct answers and reasoning for questions involving the momentum-impulse theorem.  In our prior 
study, $20\%$ of students were able to provide correct answers and reasoning after engaging in a lecture-based implementation of this same tutorial.  Although we see large gains in student 
understanding from a pre-test mean of $2\%$ for this topic, the use of the tutorials as an initial introduction to the impulse-momentum theorem appears to perform only marginally better than 
traditional instruction.

In contrast, more students had prior knowledge of the work-energy theorem as evidenced by the pre-test scores. As Figure~\ref{fig:chart} demonstrates, we find a progressive improvement in student post-test 
performance as the instruction moves closer to the ideal implementation model previously defined.  In terms of statically significant differences between post-test mean scores, we find Style 2 
(complete written answers provided) to be statistically lower ($p<0.05$) than both Style 4 (limited Socratic dialogue use) and the ideal implementation in our prior study.  In addition, the ideal 
implementation style performed significantly better ($p<0.05$) than all other styles with the exception of Style 4 ($p=0.14$). While our study lacks resolution in the intermediate styles, our results 
suggest that this tutorial's effectiveness depends significantly on the quality of the teaching when the students have prior knowledge of the material. 

The post-test performance of students who participated in Style 5 was an unexpected result. This style was designed to be similar to the ideal implementation conducted by a highly trained teaching 
assistant in our prior study. The teaching assistant who conducted Style 5 in the present study was chosen as we deemed him to be our most skilled teaching assistant in the use of Socratic dialogue 
at the time of this study. The lower than expected post-test results for this style caused us to review the audio recordings of these sessions to determine how the teaching assistant handled 
students at checkpoints.  What we found was that the teaching assistant spent an exorbitant amount of time with each group, and while the interactions may be categorized as Socratic dialogue, we 
found it difficult at times to discern a coherent thread within the conversations. The teaching assistant was not efficient in determining if student reasoning was correct, and if students needed 
guidance, the teaching assistant struggled in asking appropriate questions.  Although it appears most student difficulties were eventually addressed, some checkpoints were so lengthy and lacked 
succinct summaries, that it is quite possible students exited the checkpoint without clear understanding. On the other hand, in our prior study the teaching assistant who conducted the ideal 
implementation was highly efficient and skilled in probing student reasoning and addressing student difficulties through directed questioning. Likewise, he clearly understood the purpose of the 
tutorial activity and ended each checkpoint with a summary of these ideas.  This finding highlights the need for the inclusion of elements in our tutorial training that we had not considered 
before including discussions on efficient use of questioning and the need for summarizing ideas with students.  

\section{DISCUSSION}

It was disappointing to us that the students who participated in the computer-based implementation performed so poorly on the post-test questions.  In our calculus-based physics course, even with 
our weekly tutorial training we find that the graduate and undergraduate recitation teaching assistants implement the tutorials very differently.  A computer-based environment could provide more 
consistency in this implementation as well as cost savings due to less training and fewer teaching assistants being needed.  In addition, a survey question at the end of the computer-based 
tutorial activity indicated that students actually liked the 
computer-based implementation with half including comments about their positive experience and only one student preferring the normal recitation of the calculus-based physics course. The most 
frequently mentioned advantage of the computer-based implementation was that the students felt they left with a definitive set of answers. Unfortunately, this did not transfer over to time spent 
on the activity, and we acknowledge that the lack of time on task for students who completed the computer-based version of the tutorials may have been a key factor in the low post-test scores.  
More research is needed particularly in finding ways to motivate students to spend more time engaged with the material when presented in this format.  In addition, the computer-based checkpoints 
in this study were designed to closely match the printed tutorial materials.  More research is needed to determine what type and format of feedback is most effective in this implementation style 
and also whether or not the use of cooperative learning groups further impacts student performance under this implementation. 

The difference in trends for student post-test performance on the kinetic energy and momentum questions across the different implementation styles suggests that the level of instructional support 
is more important when students are more familiar with the material.  This may seem counter-intuitive but one of the instructional benefits of Socratic dialogue is that it can help students move 
from their current ideas to more correct or complete ideas.  Without a teaching assistant probing student understanding, it is plausible that students move through the material without ever 
considering and building upon their prior knowledge. On the other hand, as in the case of the impulse-momentum theorem of which students demonstrated little prior knowledge, the instructional 
materials themselves were adequate in developing foundational knowledge.  More research is needed to investigate each tutorial's effectiveness as an initial introduction to the material as compared 
to when students have been exposed to the ideas before. This would aid in placement of recitations topics throughout the semester such that materials could be used to their utmost potential.

Results of this study suggest improvements for our weekly tutorial training. Currently our teaching assistants spend one hour per week working together through the tutorial worksheets for the 
upcoming week. They have their reasoning checked by the course coordinator, and if time permits common student misconceptions are discussed.  Limited resources do not allow us to pair new and 
experienced teaching assistants so although we discuss the use of Socratic dialogue, most have not see it in action.  Those who learn the technique over time, however, seem to struggle with 
efficient and effective use of directed questioning as observed with the teaching assistant who conducted Style 5 in this study.  We feel that a more explicit instructional approach in our weekly 
training may result in more consistent implementation of the tutorials in our recitations.  Providing teaching assistants with a list of the specific learning outcomes for each checkpoint may 
better focus attention on what students should be learning.  Providing a set of suggested checkpoint questions similar to those found in the {\em Physics by Inquiry Instructor's Guide}\cite{PbIGuide} might improve the 
efficiency and effectiveness of even our best teaching assistants.

\section{Acknowledgements}

We wish to thank the Physics Education Group at the University of Washington, and in particular, Lillian McDermott, Peter Shaffer, Paula Heron, and MacKenzie R.~Stetzer for their valuable advice 
and support.  Partial support for this work has been provided by the National Science Foundation under grant number DUE-0126919.

\end{document}